\documentclass[preprint2,numberedappendix]{emulateapj-rtx4}
\usepackage{graphicx,natbib,times}
\usepackage{bm,url}
\graphicspath{{./fig/}{./png/}}
\usepackage{color}

\newcommand{\EQ}{\begin{equation}}
\newcommand{\EN}{\end{equation}}
\newcommand{\EQA}{\begin{eqnarray}}
\newcommand{\ENA}{\end{eqnarray}}
\newcommand{\eq}[1]{(\ref{#1})}
\newcommand{\EEq}[1]{Equation~(\ref{#1})}
\newcommand{\Eq}[1]{equation~(\ref{#1})}

\newcommand{\Eqss}[2]{equations~(\ref{#1})--(\ref{#2})}

\newcommand{\App}[1]{Appendix~\ref{#1}}
\newcommand{\Sec}[1]{Section~\ref{#1}}

\newcommand{\Fig}[1]{Figure~\ref{#1}}
\newcommand{\FFig}[1]{Figure~\ref{#1}}

\newcommand{\bra}[1]{\langle #1\rangle}

{}
{}
{}

{}
{}
{}
{}
{}
{}
{}
{}
{}
{}
{}
{}
{}
{}
{}
{}
{}
{}
{}
{}
\newcommand{\meanUU}{\overline{\mbox{\boldmath $U$}}{}}{}

{}
{}
{}

{}

{}
{}

\newcommand{\kk}{\bm{k}}

\newcommand{\BB}{\bm{B}}

\newcommand{\JJ}{\mbox{\boldmath $J$} {}}

\newcommand{\AAA}{\mbox{\boldmath $A$} {}}

\newcommand{\nab}{\mbox{\boldmath $\nabla$} {}}

\newcommand{\ii}{{\rm i}}

\newcommand{\dd}{{\rm d} {}}

\newcommand{\const}{{\rm const}  {}}

\def\la{\mathrel{\mathchoice {\vcenter{\offinterlineskip\halign{\hfil
$\displaystyle##$\hfil\cr<\cr\sim\cr}}}
{\vcenter{\offinterlineskip\halign{\hfil$\textstyle##$\hfil\cr<\cr\sim\cr}}}
{\vcenter{\offinterlineskip\halign{\hfil$\scriptstyle##$\hfil\cr<\cr\sim\cr}}}
{\vcenter{\offinterlineskip\halign{\hfil$\scriptscriptstyle##$\hfil\cr<\cr\sim\cr}}}}}
\def\ga{\mathrel{\mathchoice {\vcenter{\offinterlineskip\halign{\hfil
$\displaystyle##$\hfil\cr>\cr\sim\cr}}}
{\vcenter{\offinterlineskip\halign{\hfil$\textstyle##$\hfil\cr>\cr\sim\cr}}}
{\vcenter{\offinterlineskip\halign{\hfil$\scriptstyle##$\hfil\cr>\cr\sim\cr}}}
{\vcenter{\offinterlineskip\halign{\hfil$\scriptscriptstyle##$\hfil\cr>\cr\sim\cr}}}}}
\def\sinc{\mbox{\rm sinc}}
\def\atan{\mbox{\rm arctan}}
\def\Atan{\mbox{\rm Arctan}}

\def\Sh{\mbox{\rm Sh}}

\def\Pm{\mbox{\rm Pr}_M}
\def\calR{{\cal R}}
\def\RM{{\rm RM}}
\def\Rm{R_{\rm m}}

\def\kf{k_{\rm f}}

\def\urms{u_{\rm rms}}

\def\half{{\textstyle{1\over2}}}

\newcommand{\uG}{\,\mu{\rm G}}

\newcommand{\cm}{\,{\rm cm}}

\newcommand{\m}{\,{\rm m}}

\newcommand{\rad}{\,{\rm rad}}

\newcommand{\pc}{\,{\rm pc}}
\newcommand{\kpc}{\,{\rm kpc}}

\newcommand{\yapj}[3]{ #1, {ApJ,} {#2}, #3}

\newcommand{\yapjl}[3]{ #1, {ApJ,} {#2}, #3}

\newcommand{\yana}[3]{ #1, {A\&A,} {#2}, #3}
\newcommand{\yanas}[3]{ #1, {A\&AS,} {#2}, #3}

\newcommand{\yjetpl}[3]{ #1, {JETP Lett.,} {#2}, #3}

\newcommand{\yaraa}[3]{ #1, {ARA\&A,} {#2}, #3}

\newcommand{\ymn}[3]{ #1, {MNRAS,} {#2}, #3}

\newcommand{\ypnas}[3]{ #1, {Proc.\ Nat.\ Acad.\ Sci.,} {#2}, #3}

\newcommand{\yjour}[4]{ #1, {#2}, {#3}, #4}

\newcommand{\ybook}[3]{ #1, {#2} (#3)}
\newcommand{\yproc}[5]{ #1, in {#3}, ed.\ #4 (#5), #2}

\begin{document}
\title{
Faraday signature of magnetic helicity from reduced depolarization
}\author{
Axel Brandenburg$^{1,2}$
and Rodion Stepanov$^{3,4}$
}\affil{
$^1$Nordita, KTH Royal Institute of Technology and Stockholm University,
Roslagstullsbacken 23, 10691 Stockholm, Sweden\\
$^2$Department of Astronomy, AlbaNova University Center, Stockholm University,
10691 Stockholm, Sweden\\
$^3$Institute of Continuous Media Mechanics, Korolyov str.\ 1,
614013 Perm, Russia\\
$^4$Perm National Research Polytechnic University, Komsomolskii Av. 29, 614990 Perm, Russia
}

\date{Received 2014 January 16, accepted 2014 March 18,~ $ $Revision: 1.253 $ $}

\label{firstpage}

\begin{abstract}
Using one-dimensional models, we show
that a helical magnetic field with an appropriate sign of helicity
can compensate the Faraday depolarization
resulting from the superposition of Faraday-rotated polarization planes
from a spatially extended source.
For radio emission from a helical magnetic field, the polarization as a
function of the square of the wavelength becomes asymmetric with respect
to zero.
Mathematically speaking, the resulting emission occurs then either at
observable or at unobservable (imaginary) wavelengths.
We demonstrate that rotation measure (RM) synthesis allows for the
reconstruction of the underlying Faraday dispersion function in the
former case, but not in the latter.
The presence of positive magnetic helicity can thus be detected by observing
positive RM in highly polarized regions in the sky and
negative RM in weakly polarized regions.
Conversely, negative magnetic helicity can be detected by observing
negative RM in highly polarized regions and
positive RM in weakly polarized regions.
The simultaneous presence of two magnetic constituents with
opposite signs of helicity is shown to possess signatures that can be
quantified through polarization peaks at specific wavelengths and
the gradient of the phase of the Faraday dispersion function.
Similar polarization peaks can tentatively also be identified
for the bi-helical
magnetic fields that are generated self-consistently by a dynamo
from helically forced turbulence, even though
the magnetic energy spectrum is then continuous.
Finally, we discuss the possibility of detecting magnetic fields with
helical and non-helical
properties in external galaxies using the Square Kilometre Array.
\end{abstract}


\section{Introduction}

For many decades, polarized radio emission from external galaxies has
been used to infer the strength and structure of their magnetic field.
This emission is caused by relativistic electrons gyrating around
magnetic field lines and producing the polarized synchrotron emission.
The plane of polarization gives an indication about the electric
(and thus magnetic) field vectors at the source of emission.
The line-of-sight component of the field can be inferred through
the Faraday effect that leads to a wavelength-dependent rotation
of the plane of polarization.
The resulting change of the angle of the polarization plane over a
certain wavenumber interval gives the rotation measure (RM),
whose variation across different positions within external galaxies
gives an idea about the global structure of the magnetic fields of
these galaxies \citep{SFW86,BBMSS96,Beck05,Fle10,BW13}.

In practice, an observer will always see a superposition of different
polarization planes from different depths, which can lead to a reduction
in the degree of polarization.
Firstly, the orientation of the magnetic field changes,
causing different polarization planes at different positions.
Secondly, Faraday rotation causes the plane of polarization to rotate.
The decrease in polarized emission resulting from this superposition
is referred to as Faraday depolarization.
This was regarded as a problem that can be alleviated partially by
restricting oneself to observations at shorter wavelengths \citep{SKDU11}.
This situation has changed with the advent of new generations of
radio telescopes that can measure polarized emission over a broad
and continuous range of wavelengths.
This allows one to apply the method of \cite{Bur66} that utilizes the
wavelength-dependent depolarization to determine the distribution of
radio sources with respect to Faraday depth \citep{BB05,HBE09,Gie13,FSSB11}.
However, the interpretation of distributed magnetic fields
still remains a challenge \citep{BFSS12,BE12}.

Of particular interest to the present study is the possibility
of detecting helicity of the magnetic field.
The helicity of the magnetic field reflects the linkage of the
magnetic field \citep{Mof78}.
In the context of the large-scale magnetic field in galaxies,
one can think of the linkage between the poloidal and toroidal
magnetic field components.
Three-dimensional visualizations of these two components together,
such as Fig.~5 of \cite{DB90}, show that the magnetic field lines
describe a spiralling pattern.
Another manifestation of a helical field is the rotation of
a magnetic field vector perpendicular to the line of sight.
Determining the presence of such swirling magnetic field patterns
would be an important step toward understanding the nature of
the underlying dynamo process that is needed to achieve better
agreement between observations and theory of astrophysical dynamos.
A promising result for probing magnetic helicity in the
interstellar medium has been obtained by \cite{VS10}, who have
shown that a helical turbulent magnetic field produces a nonzero
cross-correlation of $\RM$ and the degree of polarization.
The sign of the cross-correlation coefficient permits one
to define the sign of the total magnetic helicity.
However, the theoretical background of this approach was not
clearly understood.
Subsequent attempts by \cite{JE11} and \cite{OJRE11} did not clarify
this effect either, because they excluded the effect of
Faraday depolarization from the beginning.
To explain the results of \cite{VS10}, we stress the fact that,
if the magnetic field is helical, i.e.,
the magnetic field lines spiral toward or away from the observer,
the resulting Faraday depolarization can be either enhanced or reduced,
depending on the relative signs of magnetic helicity and the line-of-sight
component of the magnetic field and thus $\RM$.
In a related paper by \cite{HF14}, this effect was used to study
the polarized intensity in selected wavelength ranges for
both signs of helicity.
The exploitation of this effect, which was first discussed by \cite{Soko98}
as an anomalous depolarization due to a twisted magnetic field,
is an important motivation behind the present paper.

While the effect of a helical magnetic field is easily understood
for simple magnetic spirals, it becomes less obvious in the case
of more complicated fields.
We are here particularly interested in helical magnetic fields consisting
of constituents that have large and small length scales with opposite signs
of magnetic helicity.
Such fields are called bi-helical and are of central importance in dynamo theory
\citep[for a review, see][]{BS05} and have also been detected in the
solar wind \citep{BSBG11} and on the solar surface \citep{ZBS14}.
There is now also some evidence for helical magnetic fields in the jets
emanating from active galactic nuclei \citep{RG12}.
We first discuss the observational signatures of singly helical fields
and turn then to the case of bi-helical magnetic fields.
Next, we discuss a method referred to as
cross-correlation analysis using magnetic field
configurations similar to those studied in the first part of the paper.
Those fields are used to mimic the effects of turbulence
consisting of randomly oriented patches with
singly helical or bi-helical fields oriented randomly in the sky.
Finally, we present preliminary results from more realistic magnetic field
configurations generated by a turbulent dynamo in the presence of shear.
We conclude with a discussion of the possibilities of detecting
helical and bi-helical magnetic fields in external galaxies using
the Square Kilometre Array.

\begin{figure}\begin{center}
\includegraphics[width=\columnwidth]{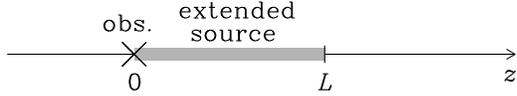}
\end{center}\caption[]{
Sketch illustrating position of source and observer.
}\label{sketch}\end{figure}

\section{Compensating depolarization}
\label{Cancelling}

The synchrotron emission of magnetized interstellar or intergalactic media
is commonly observed through its total intensity,
\begin{equation}
I(\lambda^2)= \int_0^\infty \epsilon(z,\lambda) \,\dd z,
\label{eq:tot}
\end{equation}
and through the Stokes $Q$ and $U$ parameters
combined into a complex polarization as
\begin{equation}
P(\lambda^2)\equiv Q+\ii U=p_0 \int_0^\infty
\epsilon(z,\lambda) e^{2\ii(\psi(z)+\phi(z)\lambda^2)}  \,\dd z,
\label{eq:pol}
\end{equation}
at a given point in the sky.
Here $p_0$ is the intrinsic polarization
(depending on the energy spectrum of the cosmic rays),
$\epsilon(z,\lambda)\propto n_c(z)B_\bot^\sigma(z) f(\lambda)$
is the polarized emissivity
with $\sigma\approx1.9$ being an exponent related to the spectral index
\citep{GS65}, $n_c$ is the cosmic-ray electron density,
$B_\bot$ is the strength of the magnetic field perpendicular
to the line of sight, $f(\lambda)\propto\lambda^{\sigma-1}$ is a
wavelength-dependent factor, $\psi(z)$ is the intrinsic polarization angle,
$K=0.81\m^{-2}\cm^3\uG^{-1}\pc^{-1}$ is a constant \citep{Pac70},
$\lambda$ is the wavelength,
\begin{equation}
\phi(z)=-K \int_0^z \! n_e(s) B_\|(s) \, \dd s.
\label{eq:fd}
\end{equation}
is the Faraday depth, $n_e$ is the electron density (dominated by thermal
electrons), $B_\|$ is the magnetic field along the line of sight,
and $z$ is a coordinate along the line of sight
in a Cartesian coordinate system, $(x,y,z)$.
Note that \Eq{eq:fd} implies that the Faraday depth is positive
when the mean magnetic field points toward the observer at $z=0$;
see \Fig{sketch} and \App{App} for alternative conventions
concerning \Eqss{eq:tot}{eq:fd}.
Variations across the sky are here ignored, so there is no dependence
on $x$ and $y$; see \cite{DB90}, \cite{EMB92}, \cite{BDMSS93}, and
\cite{UEB97} for
early applications to mean-field dynamos where this restriction was relaxed.
Note that $\epsilon$ also depends on $\lambda$ through a factor
$f(\lambda)$, but this term can be moved outside the integral,
so it does not constitute a principle problem
\citep{BB05,BE12}, and we shall ignore this complication here.
The {\em observed} polarization angle is
\EQ
\chi(\lambda^2)=\half\Atan(U,Q),
\EN
where $\Atan$ returns all angles in the range from $-\pi$ to $\pi$,
whose tangent yields $U/Q$.
It is not to be confused with the intrinsic polarization angle $\psi(z)$.

\begin{figure}\begin{center}
\includegraphics[width=\columnwidth]{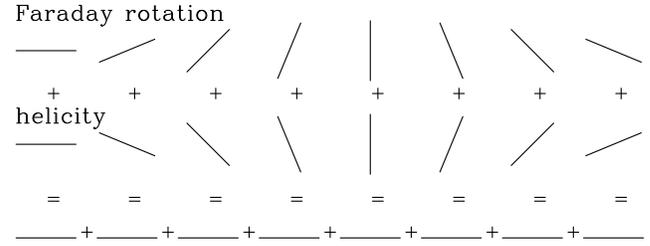}
\end{center}\caption[]{
Sketch illustrating the combined effects of Faraday rotation
and a helical magnetic field.
For a uniform magnetic field, contributions from different depths
lead to different angles of the polarization plane.
Thus, Faraday rotation alone would lead to Faraday depolarization
(sum of the phases of all contributions from the first row),
but when $\BB_\perp$ is a helical field rotating properly
about the $z$-axis (second row),
the contributions from different depths
lead to the same {\em observed} polarization angle (last row) and
Faraday depolarization is thus compensated.
}\label{cancel_depol_sketch}\end{figure}

Since $\BB$ is assumed independent of $x$ and $y$,
the divergence-free condition implies that $B_\|=B_z=\const\equiv B_{\|0}$.
While the assumed independence of $x$ and $y$ may be justified for
large-scale fields, it is certainly problematic for small-scale fields.
This will be addressed in \Sec{TurbulenceGenerated}.
We write the perpendicular magnetic field $\BB_\perp=(B_x,B_y,0)$
in complex form,
\EQ
{\cal B}(z)\equiv B_x(z)+\ii B_y(z)=B_\perp(z)\,e^{\ii\psi_B(z)}
\label{eq:B}
\EN
with its phase $\psi_B=\Atan(B_y,B_x)$.
The intrinsic polarization angle $\psi$ is related to $\psi_B$ by
\EQ
\psi=\psi_B - \pi/2.
\label{eq:psipsiB}
\EN
Here the $\pi/2$ term comes from the fact that the plane of polarization
is parallel to the electric field and perpendicular to the magnetic field
of the radio wave, which, in turn, is parallel to the ambient field $\BB_\perp$.
[Note that this term is sometimes omitted; see \cite{WSE09}
for such an example.
\cite{Soko98} included it, but dropped the resulting minus sign
after their equation (16).]
Due to the factor $2$ in the exponent of \Eq{eq:pol},
which is a consequence of the definition of the Stokes parameters
being essentially squared quantities,
the phase of the magnetic field has a $\pi$ ambiguity.
This is a serious restriction, because it means that the underlying
magnetic field cannot be determined fully without additional assumptions.

We now want to determine a condition on the structure of the magnetic field
under which the integral in \Eq{eq:pol} gives maximum contribution,
that is, for which the Faraday depolarization is minimal.
As was already shown by \cite{Soko98}, this
is the case when, for a certain value of $\lambda$,
the phase $2(\psi(z)+\phi(z)\lambda^2)$ is a constant.
For the purpose of the present discussion we assume constant values of
$B_\perp$, $n_e$, and $n_c$, denoted by
$B_{\perp0}$, $n_{e0}$, and $n_{c0}$, respectively.
Therefore, $\phi(z)=-K n_{e0} B_{\|0} z$ is linear in $z$,
and so the (half) phase under the integral in \Eq{eq:pol} is given by
\begin{equation}
\psi(z)+\phi(z)\lambda^2=\psi(z)-K n_{e0} B_{\|0}\lambda^2 z,
\label{eq:cond1}
\end{equation}
which becomes independent of $z$ and equal to a constant $\psi_0$,
giving thus maximum contribution to the integral, when
\EQ
\psi_B(z)=\psi_0-kz,
\label{chiB}
\EN
where $\psi_0$ is an arbitrary phase shift and
\EQ
k=-K n_{e0} B_{\|0}\lambda^2
\label{eq:keqn}
\EN
is the required wavenumber of the magnetic field.
A similar condition was also derived by \cite{AB11}, without however
explicitly making reference to the helical nature of the magnetic field.

\EEq{chiB} implies that we have a {\em unique solution} for the magnetic
field that gives maximum contribution to the integral in \Eq{eq:pol}
by essentially canceling the Faraday depolarization from the
$\exp(2\ii\phi\lambda^2)$ term, as illustrated in \Fig{cancel_depol_sketch}.
Inserting \Eq{chiB} into \Eq{eq:B} and assuming $B_\perp=\const$,
we have
\EQ
\BB=\left(B_{\perp0} \cos (kz-\psi_0),
-B_{\perp0} \sin(kz-\psi_0), B_{\|0}\right).
\label{eq:bb}
\EN
Such a twisted magnetic field with $\psi_B(z)\propto z$
is a Beltrami field and has been considered
by \cite{Soko98} for the demonstration of anomalous depolarization.

As motivated above, we are interested in the magnetic helicity of the field.
It is defined as $\bra{\AAA\cdot\BB}$, where angular brackets denote
volume averaging and $\AAA$ is the magnetic vector potential with
$\BB=\nab\times\AAA$ and components $\AAA=(B_x/k, B_y/k+xB_{\|0}, 0)$.
Here the linearly varying component $xB_{\|0}$ is needed to give the
constant $B_\|=B_{\|0}$,
but this contribution averages out in
the calculation of the magnetic helicity,
\begin{equation}
\bra{\AAA\cdot\BB}=k^{-1}B_{\perp0}^2.
\label{eq:hel}
\end{equation}
Another quantity of interest, which is based on the current density
$\JJ=\nab\times\BB/\mu_0$ with $\mu_0$ being the vacuum permeability,
is the current helicity, $\bra{\JJ\cdot\BB}=k B_{\perp0}^2/\mu_0$.
In the present example, it has the same sign as $\bra{\AAA\cdot\BB}$
and is positive (negative) for positive (negative) values of $k$.
Note also that $\psi_B$ decreases (increases) with $z$ when the
magnetic helicity is positive (negative).
Somewhat surprisingly, this implies that the tips of the magnetic
field vectors describe a left-handed (right-handed) spiral when
magnetic helicity is positive (negative).

For a {\em given} magnetic field, that is, prescribed $k$ and $B_{\|0}$,
$|P(\lambda^2)|$ as a function of $\lambda$ becomes maximal if \Eq{eq:keqn}
holds, that is $\lambda^2=-k/K n_{e0} B_{\|0}$.
Obviously, only $\lambda^2>0$ is observable, so only
negative  (positive) helicities can be detected
via the observation of a maximum of $|P(\lambda^2)|$
if $B_{\|0}$ is positive (negative), i.e.,
the field points away from (toward) the observer.

To give an example for typical values of the radio wavelength expected from
magnetic fields in the interstellar medium and in external galaxies,
let us take $k=2\pi/\kpc$ for the wavenumber of a field of one kpc scale,
$n_{e0}=0.03\cm^{-3}$ \citep{TC93}, and $B_{\|0}=3\uG$, then
$|P(\lambda^2)|$ peaks at $\lambda\approx30\cm$.
To probe fields with larger (smaller) length scales, one would need
shorter (longer) wavelengths of the radio emission.

\section{Faraday dispersion function}

To characterize the observational signature of a helical magnetic field,
we compute the corresponding complex polarization as a function of $\lambda^2$
using \Eq{eq:pol}.
For the purpose of further analysis the polarization can be expressed
as a Fourier integral,
\begin{equation}
P(\lambda^2)=\int_{-\infty}^{\infty} F(\phi) \, e^{2\ii\phi\lambda^2} \,\dd\phi,
\label{eq:pol2}
\end{equation}
where
\EQ
F(\phi)=f(\phi)\,e^{2\ii\psi(\phi)}
\label{eq:Fphi}
\EN
is called the Faraday dispersion function \citep{Bur66} with $f(\phi)=|F(\phi)|$.
Provided that \Eq{eq:fd} defines a strictly monotonous function $\phi(z)$,
we have $\dd\phi/\dd z\neq0$ and can change variables from $z$ to $\phi$
in \Eq{eq:pol}, and we write
\begin{equation}
f(\phi)=-p_0 \epsilon(\phi) / K n_e(\phi) B_\|(\phi),
\label{eq:fz}
\end{equation}
where the denominator is just $\dd\phi/\dd z$
resulting from the transformation from $z$ to $\phi$.
The factor $2$ in the exponent of \Eq{eq:Fphi} results in
the $\pi$ ambiguity.
It is therefore useful to characterize signatures of helical
magnetic fields directly in terms of $F(\phi)$.
This is particularly important, because there is, at least in principle,
the chance to reconstruct $F(\phi)$ from $P(\lambda^2)$ using Fourier
transformation with respect to the conjugate variable $2\lambda^2$
\citep{Bur66}.
Given the lack of any information about $P(\lambda^2)$ for $\lambda^2<0$
we define the synthesized Faraday dispersion function \citep{Bur66,BB05},
\begin{equation}
F_{\rm syn}(\phi)={1\over2\pi}
\int_{0}^{\infty} P(\lambda^2) \, e^{-2\ii\phi\lambda^2} \,\dd(2\lambda^2),
\label{eq:RMsyn}
\end{equation}
which is supposed to be a reasonable approximation of the actual $F(\phi)$,
which would be obtained if the integral in \Eq{eq:RMsyn}
were from $-\infty$ to $\infty$.

We now consider a concrete example using \Eq{eq:bb} with $k=k_1$ to construct a
magnetic field in a slab of thickness $L$ with $0\leq z<L$.
In the following, we take $|k_1|=2\pi/L$, i.e., we have
within the slab just two nodes in each of the two components of $\BB_\perp$.
Outside this range, we assume $\BB_\perp={\bm0}$,
but we keep $B_\|=B_{\|0}$ everywhere.
The Faraday depth, $\phi=-Kn_{e0}B_{\|0}z$, is a uniformly varying coordinate
and ${\cal R}\equiv\phi(L)=-Kn_{e0}B_{\|0}L$ is the equivalent intrinsic
Faraday rotation measure or simply the Faraday thickness of the slab.
Then $\epsilon(\phi)\neq0$ is the range $0\leq\phi/{\cal R}\leq1$.
For normalization purposes we introduce here the wavelength $\lambda_1$.
It is given by
\EQ
\lambda_1^2=-k_1/K n_{e0} B_{\|0}
\label{eq:l0}
\EN
and determines the peak of the modulus of the resulting complex polarization,
\EQ
P(\lambda^2)=p_0I\,\hat{P}\left({\cal R}(\lambda^2-\lambda_1^2)\right),
\label{eq:Plam2}
\EN
where
\EQ
\hat{P}(\xi)=\left.\left(1-e^{2\ii \xi}\right)\right/2\ii \xi
\label{eq:burn}
\EN
is Burn's non-dimensional depolarization function, indicated by a hat.
It applies in the absence of magnetic helicity to a uniform slab
of Faraday thickness ${\cal R}$.
Note that in our normalization, $\hat{P}(0)=-1$, where the minus sign
is a consequence of the $\pi/2$ term in \Eq{eq:psipsiB}.
Note also that $\dd\arg(\hat{P})/\dd\xi=1$, in spite of
the factor $2$ in the exponential function in \Eq{eq:burn}.

\begin{figure}\begin{center}
\includegraphics[width=\columnwidth]{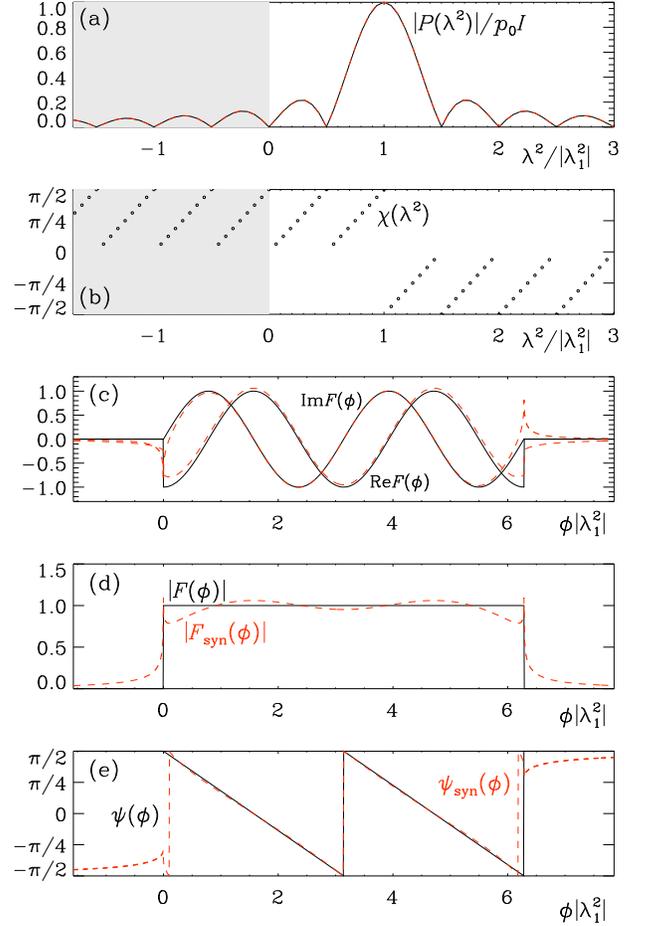}
\end{center}\caption[]{
(a) $|P(\lambda^2)|$, (b) $\chi(\lambda^2)=\arg(P)/2$,
(c) real and imaginary parts of $F(\phi)$,
(d) $|F(\phi)|$, and
(e) $\psi(\phi)$ for a
magnetic field with positive helicity $k_1>0$ and positive $\lambda_1^2>0$.
In panels (a) and (b), the unobservable range $\lambda^2<0$ is marked in gray.
In panels (c)--(e), the quantities for the synthesized Faraday dispersion function
are overplotted as red dashed lines.
}\label{helical}\end{figure}

\begin{figure}\begin{center}
\includegraphics[width=\columnwidth]{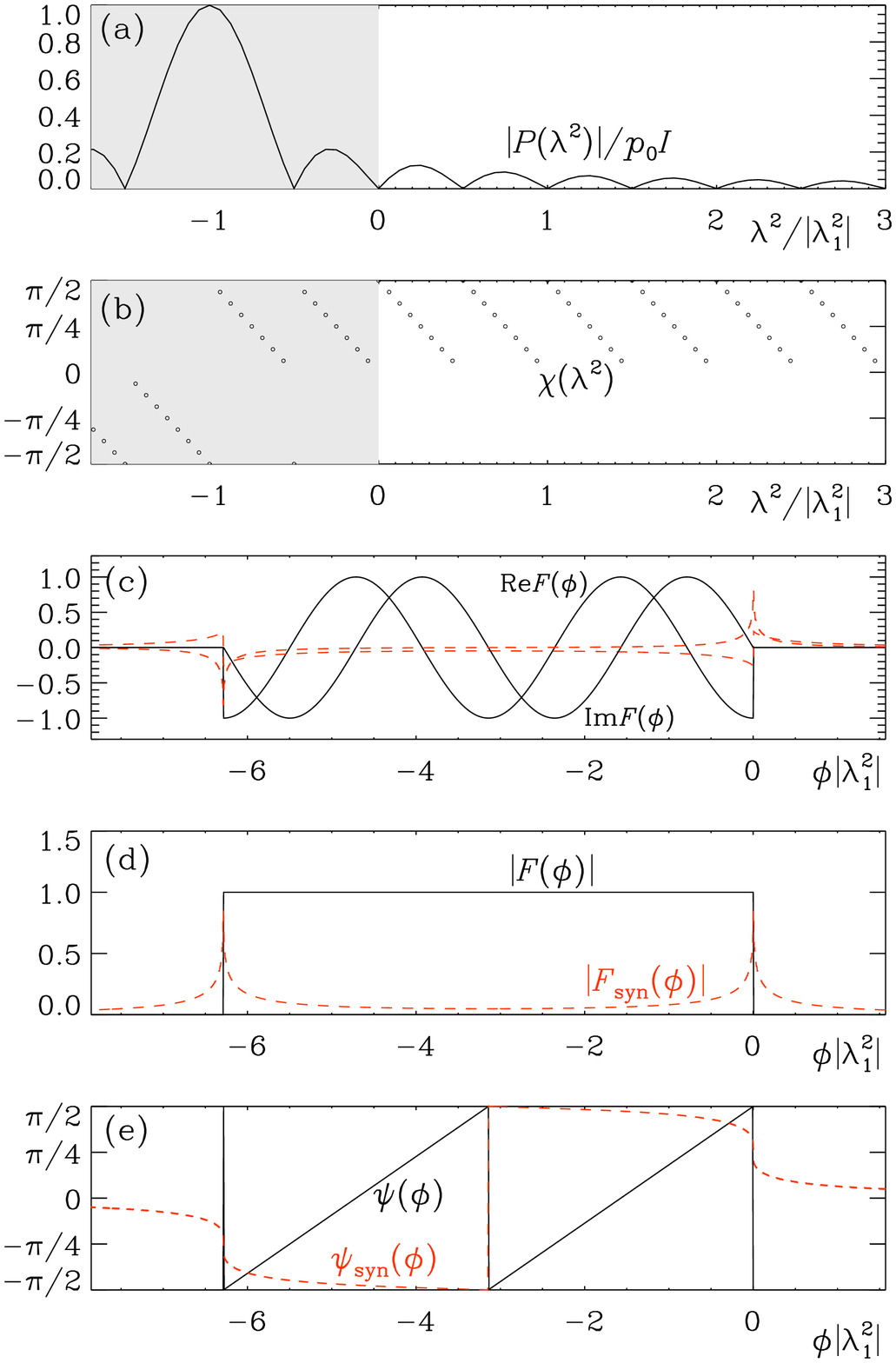}
\end{center}\caption[]{
Same as \Fig{helical}, but for $\lambda_1^2<0$,
keeping however $k_1>0$.
}\label{neghelical}\end{figure}

The resulting polarization $P(\lambda^2)$ is characterized by two
independent parameters of the magnetic field, $k_1$ and $B_{\|0}$,
which are represented by $\lambda_1^2$ and ${\cal R}$ in \Eq{eq:Plam2}.
To analyze the form of $P(\lambda^2)$,
we consider its modulus and half-phase $\chi(\lambda^2)$ and compare
the corresponding functions $F(\phi)$ and $F_{\rm syn}(\phi)$
for a helical magnetic field with positive helicity ($k_1>0$)
and different signs of $\lambda_1^2$ (\Fig{helical} for $\lambda_1^2>0$
and \Fig{neghelical} for $\lambda_1^2<0$).
We see that, as expected, $|P(\lambda^2)|$ shows a peak at
$\lambda^2=\lambda_1^2$, and the sign of $\lambda_1^2$ depends only
on that of the product of $k_1$ and $B_{\|0}$.
The polarization angle increases (decreases) with $\lambda^2$
for $k_1>0$ as shown in \Fig{helical}b (\Fig{neghelical}b).
This means that the observed rotation measure,
$\RM=\dd\chi/\dd\lambda^2$, is positive (negative).
Indeed, the case $\lambda_1^2 k_1>0$ corresponds to $\RM>0$
($B_{\|0}<0$, $\BB_\|$ toward the observer),
while $\lambda_1^2 k_1<0$ corresponds to $\RM<0$
($B_{\|0}>0$, $\BB_\|$ points away from the observer).

We note that $\RM$ does not depend on $\lambda^2$ and that
its value is half the Faraday thickness of the slab, i.e.,
$\RM={\cal R}/2$.
As mentioned above, the reason for the 1/2 factor lies in the mathematical
fact that the gradient of the phase of $\hat{P}$ in \Eq{eq:burn} is $1$
and not $2$.
It is in agreement with the interpretation that for $|F(\phi)|=\const$,
$\RM$ is the average value of $\phi$ across the source
with $0\leq\phi/{\cal R}\leq1$.

Looking at Figs.~\ref{helical}b and \ref{neghelical}b, we confirm
that at the position of the peak at $\lambda^2=\lambda_1^2$
the value of $\chi(\lambda^2)$ is $\pi/2$.
Again, this is a consequence of the $\pi/2$ term in \Eq{eq:psipsiB}
resulting from the phase shift between magnetic and electric fields
of the radio wave and the resulting effect on the plane of polarization.
Note also that $\chi(\lambda^2)$ jumps by $\pi/2$ when
$P(\lambda^2)=0$, which is the case when $\lambda^2-\lambda_1^2$
is a non-vanishing half-integer multiple of $|\lambda_1^2|$.
Unlike the jump at $\lambda^2=\lambda_1^2$ by $\pi$
because of $\pi$ ambiguity, the $\pi/2$ jumps are
physical singularities in the polarization angle as a function of $\lambda^2$.
These $\pi/2$ discontinuities were also noted by \cite{Bur66}
and are a natural consequence of decomposing a complex function with
zeros such as \Eq{eq:burn} into modulus and phase.

Since the product of $k_1$ and $\RM$ is positive in \Fig{helical},
polarized emission occurs now in the
range $0<\lambda^2<\infty$ and would therefore be observable.
As expected, the synthesized $F_{\rm syn}(\phi)$ agrees therefore
fairly well with the original $F(\phi)$; compare
the black with the red dashed lines in \Fig{helical}.
Real and imaginary parts of $F(\phi)$ and $F_{\rm syn}(\phi)$
are phase-shifted by $\pi/2$ relative to each other,
which is indicative of a helical field; see \Eq{eq:bb}.
Note also that $|F(\phi)|$ is constant and $\psi(\phi)$ is decreasing
with increasing $\phi$, as seen from \Eq{chiB}.
Again, the agreement between $F(\phi)$ and $F_{\rm syn}(\phi)$ is rather good.

If $k_1\RM<0$, the peak occurs at negative values of
$\lambda^2$ and is thus unobservable.
In that case, there would be essentially no polarized emission
and the RM-synthesized
Faraday dispersion function is very poor; see \Fig{neghelical}c--e.
A quantitative analysis of the reconstruction of the Faraday dispersion function
for different wavelength ranges and radio telescopes is given by \cite{HF14}.
The width of the polarization peaks depends on ${\cal R}$.
It is sharper for a thicker emitting region and broader for a thinner one.
In the limit of an infinitely thick slab, $P(\lambda^2)$ becomes a
$\delta$ function with no side lobes, so the remaining discrepancy between
$F(\phi)$ and $F_{\rm syn}(\phi)$ in \Fig{helical}c--e would disappear.
Perfect reconstruction of a non-helical magnetic field in a slab can
be achieved only with additional assumptions about the symmetry of the
source \citep{FSSB10}.

\section{Bi-helical magnetic fields}

In galaxies, magnetic fields are thought to be produced and maintained
by a turbulent dynamo involving a so-called $\alpha$ effect.
This leads to helical large-scale magnetic fields \citep[e.g.][]{Mof78}.
However, since magnetic helicity is an invariant in ideal
magnetohydrodynamics \citep{Wol58}, no net magnetic helicity can be produced.
Instead, a bi-helical magnetic field is generated, which
has an additional small-scale constituent of opposite magnetic helicity.
This is an idealized situation, because in reality there will be
magnetic helicity fluxes \citep{KMRS00}
that influence the local helicity balance.
Nevertheless, to study this idealized case in more detail, we consider as a simple
example the following one-dimensional, bi-helical magnetic field:
\EQ
\BB=\pmatrix{
\;\;\,B_1\cos k_1 z+B_2\cos(k_2 z+\varphi)\cr
-B_1\sin k_1 z-B_2\sin(k_2 z+\varphi)\cr B_{\|0}},
\label{bihelical}
\EN
where $k_1$ is the wavenumber of the constituent with amplitude $B_1$
$k_2$ is that of the constituent with amplitude $B_2$,
and $\varphi$ is an arbitrary phase shift between the two constituents.
The magnetic and current helicities of the total field are respectively given by
\EQ
\bra{\AAA\cdot\BB}=k_1^{-1} B_1^2+k_2^{-1} B_2^2,\;\;
\mu_0\bra{\JJ\cdot\BB}=k_1 B_1^2+k_2 B_2^2.\;\;
\EN
Thus, the field has zero magnetic helicity when
$-k_2/k_1=B_2^2/B_1^2$ and zero current helicity when
$B_2^2/B_1^2$ is $-k_1/k_2$, which is just the inverse scale ratio.
The latter situation is realized in a periodic domain after
a resistive timescale \citep{B01}, while the former is expected
to hold on short timescales \citep{FB02,BB02}.
As alluded to above, in reality there are magnetic helicity fluxes.
In practice, they tend to lead to a situation that is between
these two extreme cases \citep{BCC09}.

We emphasize that the sign of $k_i$ (with $i=1$ or 2) determines
also the sign of the helicity of the corresponding field constituent.
In the following we take $k_1>0$ and $k_2<0$ with $|k_2|>k_1$,
so the field with amplitude $B_1$ is a large-scale field with positive helicity,
and that with amplitude $B_2$ is a small-scale one with negative helicity.
This is also the situation expected to be applicable to the upper
disk plane of galaxies, i.e., where the angular velocity vector
points in the opposite direction as gravity.

\begin{figure}\begin{center}
\includegraphics[width=\columnwidth]{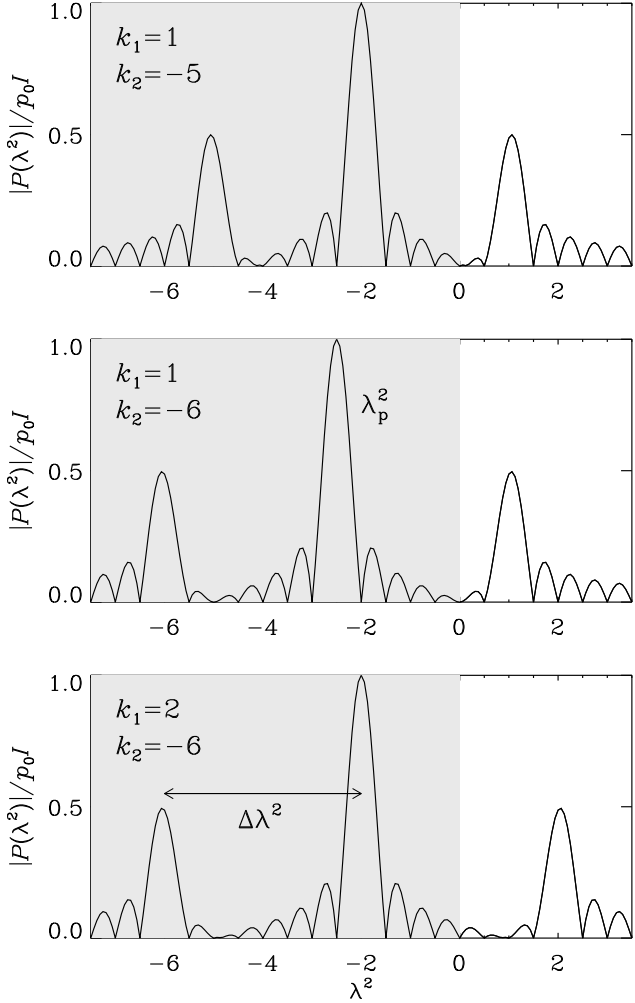}
\end{center}\caption[]{
$|P(\lambda^2)|$ for different values of $k_1$ and $k_2$
and $B_2/B_1=1$ using $\RM>0$.
The unobservable range $\lambda^2<0$ is marked in gray.
}\label{bihelical_k2}\end{figure}

\begin{figure}\begin{center}
\includegraphics[width=\columnwidth]{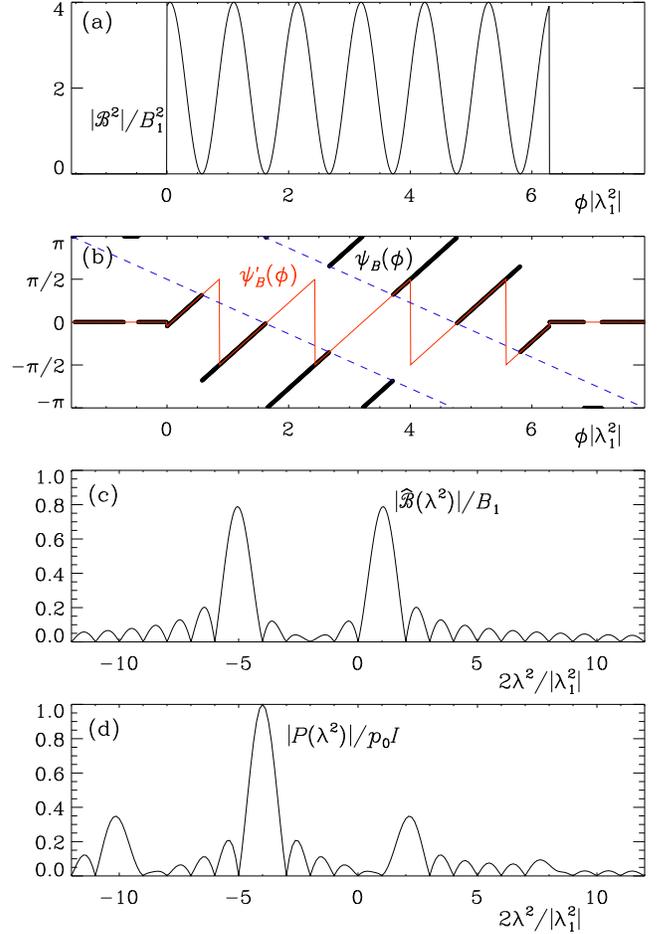}
\end{center}\caption[]{
(a) $|B|^2(\phi)$, (b) $\psi_B(\phi)$ and $\psi_B'(\phi)$,
(c) $\hat{\cal B}(2\lambda^2)$, and (d) $P(2\lambda^2)$
for a bi-helical magnetic field with $k_2/k_1=-5$
using $\RM>0$.
In panel (b), the dashed blue lines correspond to
$\pi/2-\phi|\lambda_1^2|$ and $3\pi/2-\phi|\lambda_1^2|$
and mark the points where the phase of $\psi_B(\phi)$ jumps.
}\label{dft}\end{figure}

We vary $k_1$ and $k_2$ to identify features in the results
for $P(\lambda^2)$ and $F(\phi)$
that can be related to these wavenumbers.
We define corresponding wavenumbers in Faraday space
\EQ
\lambda_i^2=-k_i/K n_{e0} B_{\|0},
\EN
which we use to define the two quantities
\EQ
\lambda_p^2=(\lambda_1^2+\lambda_2^2)/2
\quad\mbox{and}\quad
\Delta\lambda^2=(\lambda_1^2-\lambda_2^2)/2.
\EN
Note that, even though each of the two constituents of the bi-helical
field has a constant modulus, the modulus of the sum is not constant.
Instead, it is seen from the example shown in \Fig{dft} that
it varies periodically like
\EQ
|\hat{\cal B}|^2\sim\cos(2\phi\Delta \lambda^2 - \varphi).
\EN
Under the assumption that the exponent of the polarized emissivity
is $\sigma=2$, an analytic solution \Eq{eq:pol}
can be given in terms of Burn's
depolarization function \eq{eq:burn} as
\EQA
P(\lambda^2)/p_0I
&=&\epsilon_1\hat{P}\left({\cal R}(\lambda^2-\lambda_1^2)\right)\nonumber \\
&+&\epsilon_2\hat{P}\left({\cal R}(\lambda^2-\lambda_2^2)\right)\nonumber \\
&+&\epsilon_p\hat{P}\left({\cal R}(\lambda^2-\lambda_p^2)\right),
\label{eq:anal3}
\ENA
where $\epsilon_1=B_1^2/B_*^2$, $\epsilon_2=B_2^2/B_*^2$, and
$\epsilon_p=2B_1B_2/B_*^2$, with
$B_*^2=B_1^2+B_2^2+2B_1B_2\,\sinc(2\Delta\lambda^2)$,
are normalization constants.
There are three peaks of $P(\lambda^2)$:
two peaks are located at $\lambda_1^2$ and $\lambda_2^2$
and a third one appears at $\lambda_p^2$.
They are shown in \Fig{bihelical_k2} for the case $B_2/B_1=1$.
As is clear from \Eq{eq:anal3}, the separation between
adjacent peaks is given by $|\Delta\lambda^2|$.
This solution is independent of the phase shift $\varphi$
between the two constituents.

To understand the signatures of a bi-helical magnetic field
in the Faraday dispersion function,
let us recall that the wavenumbers of each of the two
constituents contribute to the gradient $\dd\psi/\dd\phi$.
It is therefore plausible that in the case $B_1=B_2$ the result is just
the average of the two, i.e.,
\EQ
\dd\psi/\dd\phi=-\lambda_p^2.
\EN
This property of $\dd\psi/\dd\phi$ is preserved
regardless of the $\pi$ ambiguity.
To demonstrate this, we compare in \Fig{dft} both $\psi_B\equiv\Atan(B_y,B_x)$
(all angles in the range from $-\pi$ to $\pi$ that yield $B_y/B_x$) and
$\psi_B'=\atan(B_y/B_x)$,
which is confined to the range from $-\pi/2$ to $\pi/2$.
As stated in \Sec{Cancelling}, $\dd\psi_B/\dd\phi$ is negative
when the product $k B_{\|0}$ is positive.
This is indeed in agreement with \Fig{dft}.

Interestingly, $\psi_B'$ is simpler than $\psi_B$ in that the
former has no phase jumps other than those required for
$\psi_B'$ to remain in the range from $-\pi/2$ to $\pi/2$.
By contrast, $\psi_B$ shows phase jumps by $\pi$ at {\em all} locations
where $|{\cal B}|$ vanishes; compare Figs.~\ref{dft}(a) and (b).
Ignoring these phase jumps, i.e., reconstructing the field from
$|{\cal B}|$ and $\psi_B'$, instead of $\psi_B$, would render the
underlying magnetic field discontinuous.

Our statements can be confirmed by evaluating \Eq{eq:anal3}
or by computing numerically examples for different combinations
of $k_1$ and $k_2$; see also \Fig{bihelical_k2}.
Thus, we can summarize that a bi-helical magnetic field with wavenumbers
$k_1$ and $k_2$ results in a clear signature in the Faraday dispersion
function in that the frequency of its modulus is given by $2\Delta \lambda^2$
(\Fig{dft}a), while indeed $\dd\psi/\dd\phi=-\lambda_p^2$ (\Fig{dft}b).

To appreciate the features of a bi-helical magnetic field in the
complex polarization $P$, let us note that a Fourier transformation
of the complex function ${\cal B}$, defined in \Eq{eq:B} and
now applied to the bi-helical field defined in \Eq{bihelical},
would produce peaks at wavenumbers $k_1$ and $k_2$.
However, in the Fourier transformation defined through \Eq{eq:pol2},
wavenumbers correspond to the Fourier variable $2\lambda^2$.
Thus, if the Faraday dispersion function was given by ${\cal B}(\phi)$
the corresponding Fourier transform $\hat{\cal B}(2\lambda^2)$ shows
peaks at $2\lambda^2/\lambda_1^2=1$ and $k_2/k_1=-5$; see \Fig{dft}c.
In reality, the Faraday dispersion function is given by ${\cal B}^2$
(assuming here $\sigma=2$).
A Fourier transformation of such a squared function has a peak at
$k_1+k_2$ and side lobes at $k_1+k_2\pm|k_1-k_2|=2k_1$ or $2k_2$.
Thus, the corresponding Fourier transform, which we can now call
$P(2\lambda^2)$, has peaks at $2\lambda^2/\lambda_1^2=2$ and
$2k_2/k_1=-10$, together with a larger one in between; see \Fig{dft}d.

\begin{figure*}\begin{center}
\includegraphics[width=\textwidth]{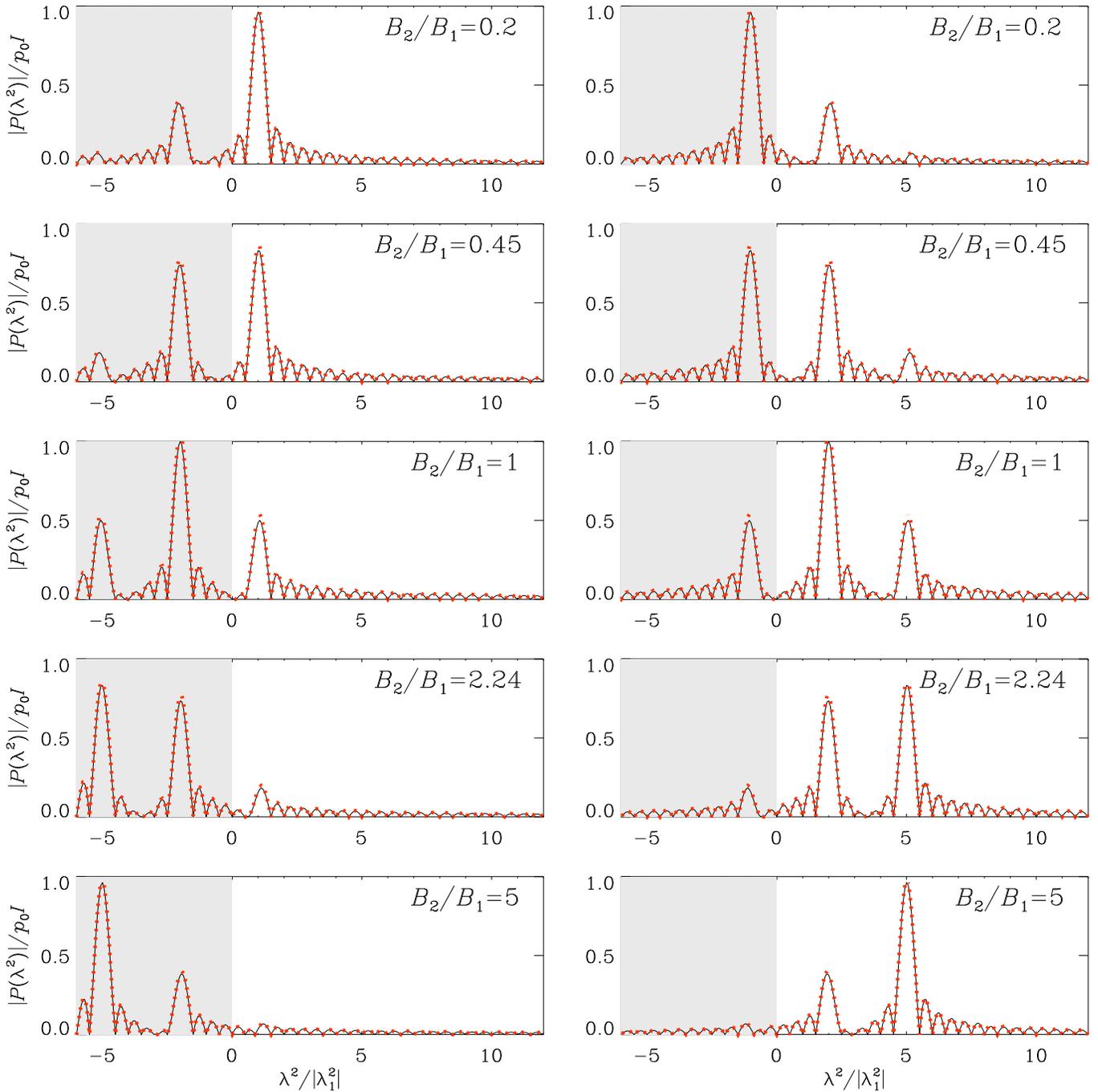}
\end{center}\caption[]{
Dependence of $|P(\lambda^2)|$ for different values of $B_1/B_2$
and $k_2/k_1=-5$ using $\RM>0$ (left column) and $\RM<0$ (right column).
The region with $\lambda^2<0$ is marked in gray.
The analytic solutions with $\sigma=2$ are shown as red dotted lines,
while the numerical one for $\sigma=1.9$ is shown as a black solid line.
For $B_2/B_1=0.45$ in the second row we have $\bra{\JJ\cdot\BB}=0$
while for $2.24$ in the fourth row we have $\bra{\AAA\cdot\BB}=0$.
}\label{bihelical_B2_both}\end{figure*}

The above considerations assume that the amplitudes of the two
constituents are approximately equal.
When $B_2/B_1$ is either very small or very large, the type of the
resulting polarization signal will be determined by the dominating
one of the two constituents.
\FFig{bihelical_B2_both} confirms that the peak at $\lambda^2=\lambda_p^2$
diminishes when $B_2/B_1$
becomes either much larger than unity or much smaller than unity.
Not surprisingly, a peak at $\lambda^2=\lambda_2^2$ begins to emerge
when $B_2$ becomes large (bottom panels of \Fig{bihelical_B2_both}),
and one at $\lambda^2=\lambda_1^2$ emerges when $B_1$ becomes large
(top panels of \Fig{bihelical_B2_both}).
In the latter case, however, most of the polarized emission occurs
formally for $\lambda^2<0$.

\FFig{bihelical_B2_both} suggests that two of the peaks have a similar height
when $\bra{\JJ\cdot\BB}=0$ (second row of \Fig{bihelical_B2_both}) or
when $\bra{\AAA\cdot\BB}=0$ (fourth row of \Fig{bihelical_B2_both}).
While this is not a general result, there is, however, a tendency
for those two peaks to survive even in the limits of very large
or very small ratios of $|k_1/k_2|$.

Our considerations of helical and bi-helical magnetic fields
have shown that the distributions of $P(\lambda^2)$ are
{\em asymmetric} with respect to $\lambda=0$.
This underlines again that the reconstruction of missing data for
negative values of $\lambda^2$ from symmetry arguments,
e.g., that $P(-\lambda^2)=P^*(\lambda^2)$,
would be impossible when the magnetic field is helical and the helicity
is of unsuitable sign (i.e., $k_1\RM<0$) for a given sign of $\RM$.
This is because the phase of the Faraday dispersion function shows
then significant
dependence on Faraday depth, so the term $\psi(z)$ cannot
be pulled outside the integral of \Eq{eq:pol},
which is a critical assumption often made in this connection
\citep{Bur66}.

It is remarkable that in all cases with helical magnetic fields,
there is a particular value $\lambda^2$ for which the polarization approaches
the maximum value of $|P|/p_0I=1$.
Depending on the relative strengths of $B_1$ and $B_2$, this peak
can be either at $\lambda^2=\lambda^2_1$, $\lambda^2_2$, or at
$\lambda_p^2=(\lambda^2_1+\lambda^2_2)/2$; see \Fig{bihelical_B2_both}
and \Eq{eq:anal3}.

\section{Cross-correlation analysis of $|P|$ versus $\RM$}

Our present investigations have implications that help understand
earlier work in the field.
Recent surveys
of polarized emission in the interstellar medium have provided continuous
distributions of $Q$ and $U$ on the sky for certain ranges of wavelengths.
Due to finite beam size, only a small number of independent lines of sight
are available for analysis.
Probing magnetic helicity with a cross-correlation analysis between $\RM$
and the polarization degree ${\cal P}\equiv|P|/p_0I$
had been suggested by \cite{VS10} using simulated data.
While the numerical demonstration of the method was convincing,
no theoretical proof or explanation had been available yet.

\begin{figure}\begin{center}
\includegraphics[width=\columnwidth]{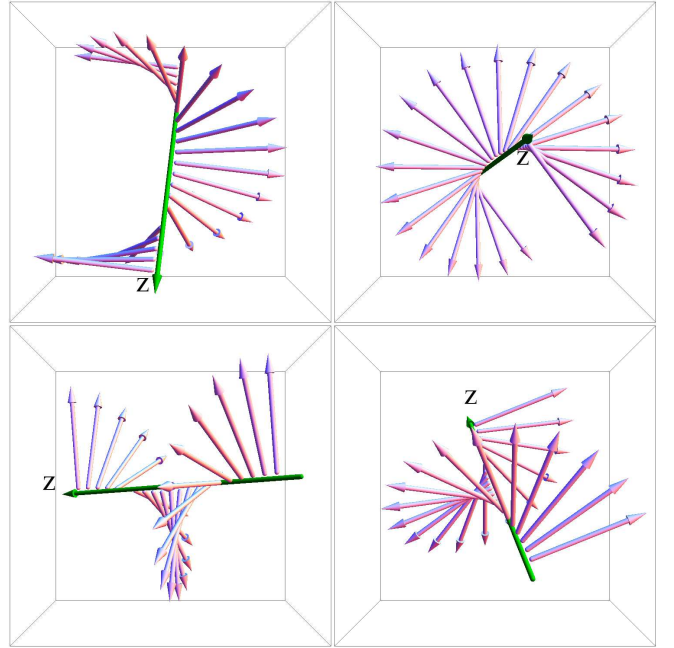}
\end{center}\caption[]{
Set of cells each with a singly helical magnetic field of positive helicity.
The tips of the vectors describe a left-handed spiral.
}\label{fig}
\end{figure}

To study this idea further, we imagine turbulence being approximated
by a set of cells possessing locally
a homogeneous helical magnetic field as in \Eq{eq:bb}.
The dominating scale of the turbulence can be attributed to
the size of the cells.
The direction of each helix is taken to be random, but for a large number
of cells there are always some
for which it is almost parallel to the
line of sight (top right panel of \Fig{fig}).
Only such cells are considered in the following.
In \cite{VS10}, the cross-correlation coefficient between synthetic maps of $\RM$ and the
polarization degree ${\cal P}$ was found to be positive (negative)
when the total magnetic helicity in the domain was
prevailingly positive (negative).
Since the direction of $B_{\|}$ is random,
the average value of $\RM$ over all cells is zero.
Then the cross-correlation coefficient is determined by the average value
of the product $\RM \,{\cal P}$, which can be considered as a weighted average
of $\RM$ with the weight ${\cal P}$.
Having in mind \Eq{eq:hel}, we recall that the maximum polarization
corresponds to cells with positive helicity and positive $\RM$ or,
alternatively, negative helicity and negative $\RM$.
Minimum polarization comes from cells with opposite sign
of helicity and $\RM$.
Thus, if the number of cells with positive and negative helicity
is about the same, then positive and negative $\RM$s are weighted equally
and the cross-correlation is zero.
If the cells with positive (negative) helicity are dominant,
then $\bra{\RM \,{\cal P}}$ is positive (negative).

\begin{figure}\begin{center}
\includegraphics[width=\columnwidth]{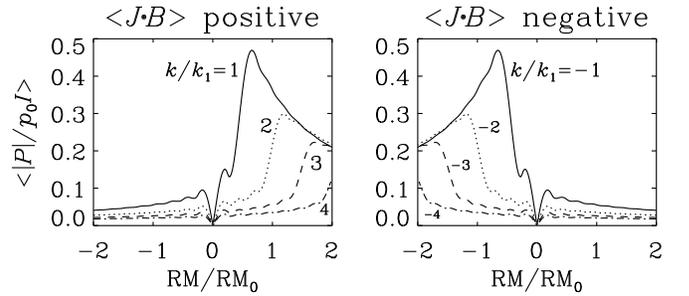}
\end{center}\caption[]{
Dependence of $\bra{P/p_0I}$ on $\RM$ for different wavenumbers
$k$ (relative to a reference wavenumber $k_1$)
and cases with positive and negative current helicities
(positive and negative values of $k$)
using an average over $0<\lambda^2/\lambda_1^2\leq10$.
}\label{panal_all2}\end{figure}

In the following, another test is suggested for the cross-correlation diagnostics.
We consider the averaged polarization $\bra{|P|/p_0I}$
by averaging over $\lambda^2$, using however only one cell.
In \Fig{panal_all2} we show first the dependence of $\bra{|P|/p_0I}$ on $\RM$
for different wavenumbers using a singly helical magnetic field.
Here we have averaged over wavelengths in the range
$0<\lambda^2\leq\lambda_1^2$.
We see that, for positive (negative) helicities, the averaged
polarization is largest for positive (negative) values of $\RM$.

\begin{figure}\begin{center}
\includegraphics[width=\columnwidth]{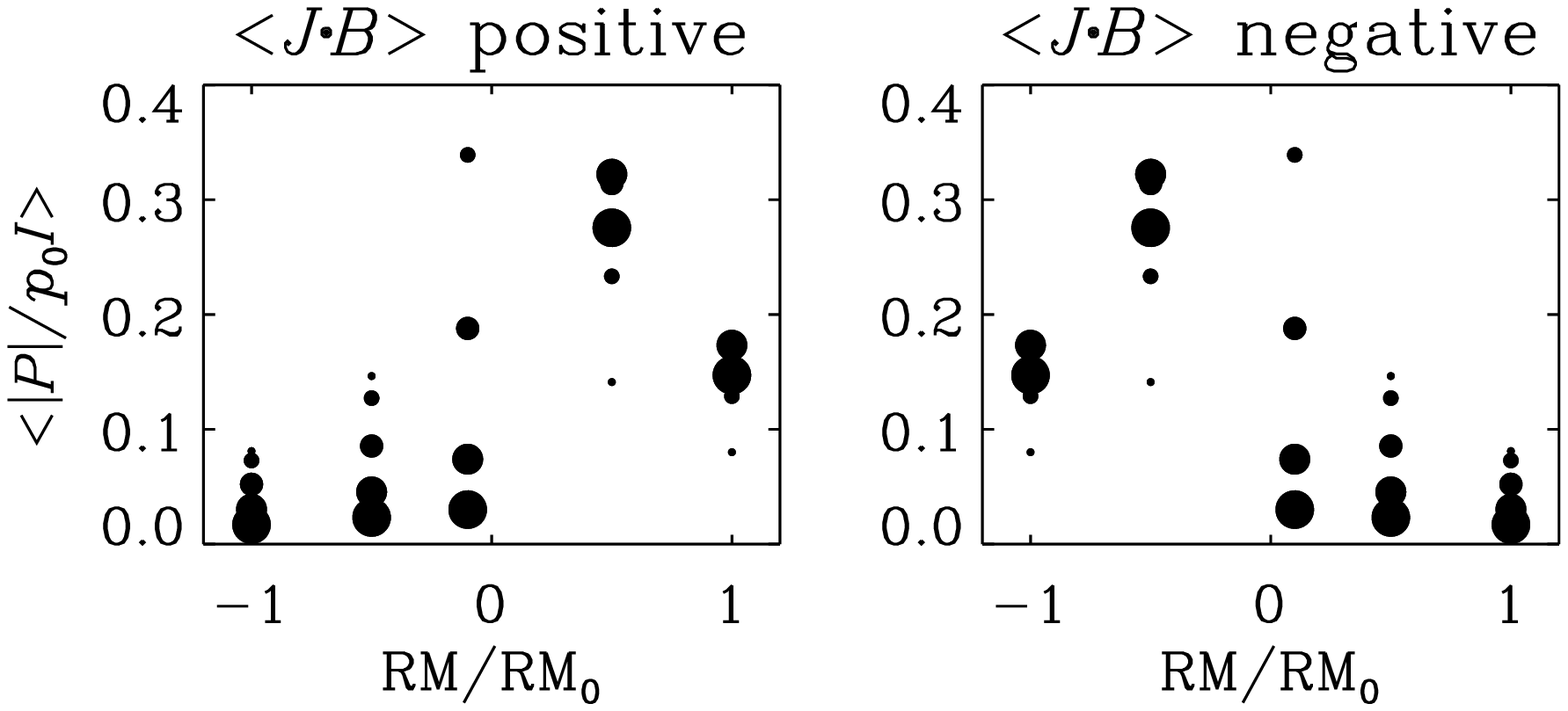}
\end{center}\caption[]{
Correlation between $\RM$ and $\bra{P/p_0I}$ for cases
with positive and negative current helicities.
The size of the symbols reflects the value of $B_2/B_1$
in \Fig{bihelical_B2_both}.
}\label{pres}\end{figure}

Next, in \Fig{pres} we show correlation plots using data from
\Fig{bihelical_B2_both} for the case of a bi-helical field,
where we take the average value of
$|P(\lambda^2)|/p_0I$ for $0<\lambda^2/\lambda_1^2\leq10$.
We also compute the corresponding results for 1/2 and 1/10
of the reference value of $\RM$, namely $\RM/\RM_0=1$, 0.5, and 0.1,
where $\RM_0\lambda_1^2=\pi$.
In the cases shown in \Fig{bihelical_B2_both}, the current helicity
$\bra{\JJ\cdot\BB}$ is negative, so the resulting polarized emission
is small for positive values of $\RM$, but large for negative values
of $\RM$.
This results in a negative correlation (see right-hand panel of
\Fig{pres}), as expected from the analysis of \cite{VS10}.
Conversely, when we change the signs of $k_1$ and $k_2$,
which corresponds to positive current helicity,
the correlation is positive.
Thus, our present results support the findings of \cite{VS10}
at a qualitative level and demonstrate, furthermore, that for bi-helical
magnetic fields their method is more sensitive to current helicity than
to magnetic helicity, which has the opposite sign in the example
considered in \Fig{pres}.

\begin{figure*}\begin{center}
\includegraphics[width=\textwidth]{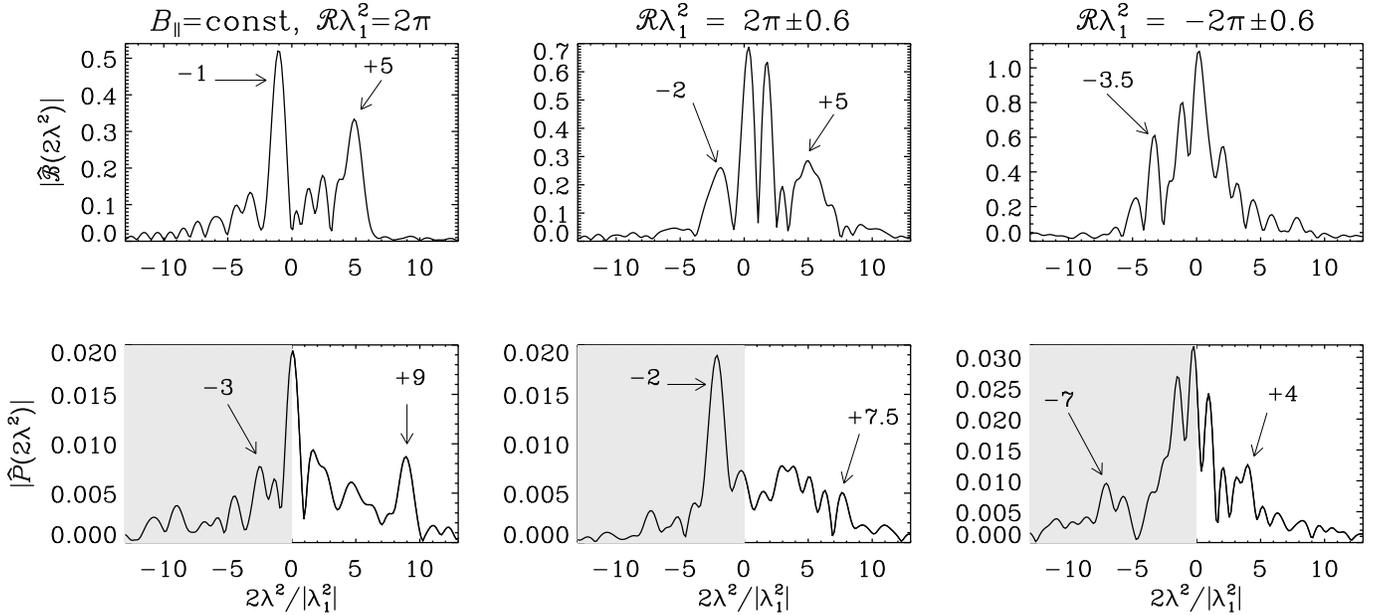}
\end{center}\caption[]{
$\hat{\cal B}(2\lambda^2)$ (upper row) and $P(2\lambda^2)$ (lower row)
for turbulence-generated magnetic fields with $k_2/|k_1|=5$,
ignoring line-of-sight variations of $B_\|$ (left column), and
including variations, shown in regions where $\calR$ is positive
(middle column) and negative (right column).
The arrows with numbers indicate particular peaks that are discussed
in the text.
}\label{pchib1_dft3}\end{figure*}

\section{Turbulence-generated magnetic fields}
\label{TurbulenceGenerated}

In this paper we have analyzed an extremely simple model of
astrophysical magnetic fields.
One potential problem is the fact that the actual magnetic field
possesses not just two scales, but there is a continuous spectrum of scales.
The other problem is that the line-of-sight magnetic field is not constant,
so $\phi(z)$ becomes nonlinear and is different for each line of sight.
To assess how the results from our idealized models are affected
by these issues, we now analyze a snapshot from a turbulence simulation
exhibiting large-scale dynamo action.

In our model, turbulence is driven through helical forcing,
as was also done in \cite{B01}, where the forcing acts only
in a narrow band of wavenumbers with an average value $\kf$ that is
five times larger than the smallest wavenumber that fits into the
computational domain, $|k_1|$.
Thus, $k_2/|k_1|=5$.
The resulting kinetic energy spectrum is, however, continuous
for $k>\kf$ and extends until the dissipative
cutoff wavenumber, whose value depends on the Reynolds number;
see Fig.~1b of \cite{BSS12} for a higher resolution simulation.
To model the effects of a significant line-of-sight magnetic field
in a physically meaningful way, we include shear.
Our model is thus similar to that of \cite{KB09}, where dynamo waves
are found to travel in the span-wise direction.
The boundary conditions are (shearing) periodic and the kinetic helicity
has the same sign throughout the computational domain, so there is no equator
in this model.

Our simulation has been carried out using the {\sc Pencil
Code}\footnote{http://pencil-code.googlecode.com/} with a resolution
of $192^3$ mesh points and is characterized by the magnetic Reynolds
and Prandtl numbers, $\Rm\equiv\urms/\eta\kf=120$ and
$\Pm\equiv\nu/\eta=1$, respectively, as well as
the shear parameter $\Sh=S/\urms\kf=0.16$.
Here $\urms$ is the rms velocity of the turbulence,
$\eta$ is the magnetic diffusivity, $\nu$ is the kinematic viscosity,
and $S=|\nab\meanUU|$ is the shear rate of the mean flow $\meanUU$.

It turns out that the nonlinearity of $\phi(z)$ is a much more serious
problem than the existence of a continuous spectrum of scales.
To demonstrate this, we begin with the best-case scenario assuming
$B_\|=\const$, so $\phi(z)$ is linear in $z$.
As in \Fig{dft}, we consider first the complex variable ${\cal B}$,
which characterizes the perpendicular magnetic field component
in the projected plane of the sky; see left column of \Fig{pchib1_dft3}.
Its Fourier transform along the line of sight, $\hat{\cal B}(2\lambda^2)$,
averaged over all points in the plane, shows clearly the small-scale
magnetic field with positive helicity at $2\lambda^2/|\lambda_1^2|=+5$
and the large-scale magnetic field with negative magnetic helicity
at $2\lambda^2/|\lambda_1^2|=-1$, corresponding to the lowest wavenumber
of the domain.
For $B_\|=\const$ and $\sigma=2$, we can
compute $|P(2\lambda^2)|$ as the Fourier transform of ${\cal B}^2$.
Its average over all points in the plane shows peaks at
$2\lambda^2/|\lambda_1^2|=-3$ (which is slightly
lower than the expected value $-2$)
and at $+9$ (which is slightly below the expected value of $+10$).
Thus, we may tentatively conclude that the presence of a continuous spectrum
of scales in the magnetic field has a less serious effect
on the polarization peaks than the nonlinearity of $\phi$
that will be discussed next.
There is, however, a peak at $\lambda^2=0$, which we have not seen
in the two-scale model.
A more detailed inspection shows that the overall depolarization is
generally rather strong when the field is turbulent.
This weakens the compensation of depolarization by helicity (\Sec{Cancelling}),
leaving behind the finite polarization at $\lambda^2=0$ due to
the contribution of a mean $\BB_\perp$ along the line of sight.
We have verified that the removal of a mean $\BB_\perp$ by replacing
$\BB_\perp\to\BB_\perp-\bra{\BB_\perp}_\|$ can reduce the peak
at $\lambda^2=0$ in most cases.

Next, we consider the effect of the nonlinearity of $\phi(z)$.
It results in regions in the plane of the sky where $\calR$ is now
either positive or negative.
Therefore, we present the results for $\hat{\cal B}$ and $P$ by averaging
over only those points where $\calR\lambda_1^2$ is in a certain interval
($2\pi\pm0.6$ and $-2\pi\pm0.6$; which is the case for about 6\%
of all lines of sight); see middle and last columns of \Fig{pchib1_dft3}.
In those points the rms value of the mean magnetic field is
about 2.5 times larger than that of the fluctuating field.
The resulting spectrum still shows some of the characteristic peaks,
but those corresponding to the large-scale field now occur at longer
wavelengths ($-2$ or $+4$ for $\calR\gtrless0$) and those corresponding to the
small-scale field at shorter wavelengths ($+7.5$ or $-7$ for $\calR\gtrless0$).
Thus, the overall result is much less clear than in the idealized model,
but some basic features of a bi-helical field can still be identified.

\begin{figure}\begin{center}
\includegraphics[width=\columnwidth]{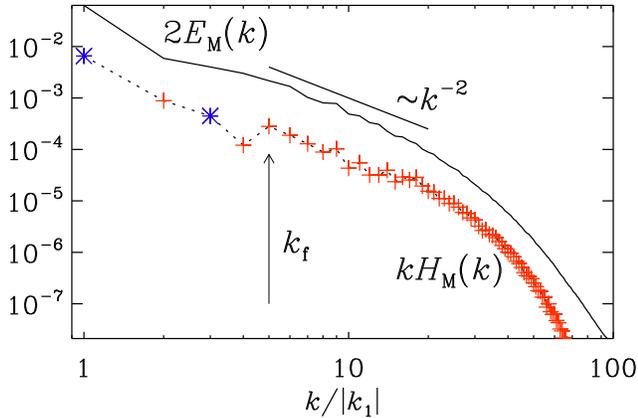}
\end{center}\caption[]{
Three-dimensional magnetic energy and rescaled helicity spectra for the
snapshot analyzed in \Fig{pchib1_dft3}.
The red plus signs indicate positive helicity and the blue asterisks
negative helicity.
}\label{ppower1}\end{figure}

In \Fig{ppower1} we show the three-dimensional magnetic energy
and helicity spectra, $E_M(k)$ and $H_M(k)$, respectively.
These spectra are normalized such that
$\int E_M(k)\,\dd k=\half\bra{\BB^2}$ and
$\int H_M(k)\,\dd k=\bra{\AAA\cdot\BB}$.
The relative magnetic helicity is defined as $r_M=k H_M(k)/2E_M(k)$,
whose modulus is between $-1$ and $+1$ \citep{Mof78,BS05}.
As expected, the field is bi-helical with negative magnetic helicity
at $k=|k_1|$ and a positive one at $k=\kf=5|k_1|$, but $r_M$ is only
about $\mp0.1$, respectively.
Contributions from $k>\kf$ are not expected to be important
because of the rapid decline of spectral power proportional to $k^{-2}$.
However, unlike the case without shear \citep[Fig.~1b of][]{BSS12},
there is no clear separation of scales and the local peak at
$k=\kf$ is barely noticeable.

Based on the results of this section, we can conclude that the reason
for the departure of $|P(\lambda^2)|$ from the ideal case is partly
the low degree of relative magnetic helicity.
However, another important reason is the occurrence of a polarization
peak at zero wavelength.
It can interfere with the other peaks and thereby contaminate
the polarization signal also at other wavelengths.

\section{Conclusions}

Our present investigations have shown that a helical magnetic field
with a suitable sign of helicity can compensate Faraday depolarization and
shift the polarized emission into the observable range.
In practice, the magnetic field has contributions from a superposition
of magnetic fields with different wavenumbers and helicities.
For bi-helical magnetic fields,
the bulk of the polarized emission is shifted to wavelengths
whose value depends on the average wavenumber of the magnetic field.
Thus, even though one of the two constituents in isolation might not be
detectable (see, e.g., the top right panel of \Fig{bihelical_B2_both}),
it could become observable because the signature of its presence
would have been carried into the observable range
(rows 3--5 on the right of \Fig{bihelical_B2_both}).
However, it is equally well possible that most of the polarized emission
would have been shifted out of the observable range
(lower panels on the left of \Fig{bihelical_B2_both}).
In that case, very little polarized emission can be expected.

When a galaxy is viewed edge-on, one can expect that its
toroidal magnetic field can provide the line-of-sight component
needed to detect helicity of field vectors in the perpendicular components.
Dynamo theory predicts that this toroidal field has the same
orientation above and below the midplane \citep{BBMSS96}.
However, the magnetic helicities of both large-scale and small-scale
fields would change sign about the equatorial plane.
Thus, it is conceivable that signatures of bi-helical magnetic fields
would be detectable on only one of the two sides around the midplane
for a fixed direction of $B_\|$.
For edge-on galaxies, this would correspond to two opposite quadrants
of detectability in the projection on the sky.

Radio emission at long (short) wavelengths would give information
about magnetic fields with large (small) wavenumbers, corresponding
to small (large) length scales.
In galaxies, the typical scales of large-scale and small-scale
magnetic fields are $1\kpc$ and $\la0.1\kpc$, respectively.
The corresponding wavenumbers are $6\kpc^{-1}$ and $\ga60\kpc^{-1}$,
respectively.
With the numbers given at the end of \Sec{Cancelling}, the corresponding
radio wavelengths would be $\lambda_1=30\cm$ for the large-scale field
and $\lambda_2\ga1\m$ for the small-scale field; see \cite{HF14} for
more detailed estimates.
However, to resolve $P(\lambda^2)$ sufficiently well, it is necessary
to sample both shorter and longer wavelengths.
With the Square Kilometre Array, we expect to obtain polarization
measurements in the range from $2\cm$ to $6\m$.
With our estimate of $\lambda_1=30\cm$ for $k_1=6\kpc^{-1}$,
this would allow access to $\lambda^2/\lambda_1^2$ from 0.004 to 400,
corresponding to $k$ from $0.03\kpc^{-1}$ to $2400\kpc^{-1}$
($=2.4\pc^{-1}$) and thus spatial scales between $240\kpc$ and $3\pc$.
This would well be compatible with the requirements for the
detection of magnetic fields with helical and bi-helical properties
in external galaxies by a safe margin.
On the other hand, our estimates are still quite rough and not yet
based on actual turbulent dynamo simulations such as those of \cite{Gre08}.
For example, if the value of $n_e B_\|$ was smaller by a factor
of 10 or more, this would easily necessitate access to the
longer wavelength range.
More importantly, contributions of the small-scale magnetic field
to $B_\|$ would substantially weaken the dependence of polarization
on $\lambda^2$.
Preliminary turbulence simulations suggest that this is indeed the case,
although the basic features of the bi-helical magnetic field resulting
from a turbulent dynamo can still be identified even then.
Further studies of such more
realistic models will be needed to assess the critical value
of small-scale contributions that can still be tolerated in $B_\|$.
There are also constraints from limited sensitivity and confusion
of the signal due to turbulence affecting all spatial scales
corresponding to radio wavelengths above $\lambda_2$.
One might speculate that this might have a tendency of reducing the
radio wavelength of the peak resulting from the small-scale magnetic field
and enhancing the wavelength of the peak resulting from the large-scale field.

An alternative diagnostic for the presence and
sign of helicity in the case of a continuous spectrum of scales
is the cross-correlation analysis of \cite{VS10}.
Surveys of polarized emission from diffuse turbulent sources in the
magnetized interstellar medium could provide appropriate data.
The presence of positive current helicity can be detected
by observing positive RM in highly polarized regions in the sky and
negative RM in weakly polarized regions.
Conversely, negative magnetic helicity can be detected by observing
negative RM in highly polarized regions and
positive RM in weakly polarized regions.
The cross-correlation coefficient between the degree of polarization and RM
provides the relevant statistical diagnostics.
Alternatively, polarization can be used instead of polarization degree.
However, in that case a nonzero cross-correlation coefficient would be
harder to distinguish.

Other possible targets where one can search for helical magnetic fields
include the ejecta from active galactic nuclei, where evidence for
swirling magnetic fields has been presented recently \citep{RG12},
and supernova remnants, which can accelerate cosmic-ray protons across
the shock, leading to a current with a component parallel to the magnetic
field, which drives current helicity and an $\alpha$ effect \citep{RKBE12}.
The typical radio wavelengths associated with helical magnetic fields
can be estimated based on their estimated Faraday depths.
For the supernova remnant G296.5+10.0, \cite{Har10} found
regions with $\RM=-14\rad\m^{-2}$ and $28\rad\m^{-2}$,
corresponding to $\lambda=\sqrt{N/2\pi\RM}\approx8$--$10\cm$,
where we have assumed $N=2$ for the number of nodes in the slab.
However, $\RM$ can show large variations and values of
$130\rad\m^{-2}$ have been suggested for G152.4-2.1 \citep{Fos13},
which would correspond to $\lambda=3.4\cm$.
This would still be within the limits of what is feasible with
present and future facilities.

\section*{Acknowledgements}

We thank Oliver Gressel for organizing the Nordita meeting on
Galactic Magnetism in the Era of LOFAR and SKA for providing
an inspiring atmosphere, where the present work was started.
We also acknowledge discussions with Cathy Horellou and Andrew Fletcher
and thank them for sharing their related results with us.
We thank Rainer Beck, Matthias Rheinhardt, Anvar Shukurov,
Kandaswamy Subramanian, and the referee for useful comments and
detailed suggestions that have led to improvements of the manuscript.
Financial support from the European Research Council under the AstroDyn
Research Project 227952, the Swedish Research Council under the grants
621-2011-5076 and 2012-5797, as well as the Research Council of Norway
under the FRINATEK grant 231444 are gratefully acknowledged.
R.S.\ acknowledges support from
the grant YD-520.2013.2 of the Council of the President of
the Russian Federation and benefitted from the International Research
Group Program supported by the Perm region government.
We acknowledge the allocation of computing resources provided by the
Swedish National Allocations Committee at the Center for
Parallel Computers at the Royal Institute of Technology in
Stockholm and the National Supercomputer Centers in Link\"oping, the High
Performance Computing Center North in Ume\aa,
the Nordic High Performance Computing Center in Reykjavik, and the
supercomputer URAN of the Institute of Mathematics and Mechanics UrB RAS.

\appendix

\section{Concerning equations (1)--(3)}
\label{App}

The purpose of this appendix is to clarify alternative definitions
of \Eqss{eq:tot}{eq:fd} in the literature.
They are related to the position of the observer,
the direction of the line-of-sight magnetic field,
and the sign of the Faraday depth.
We discuss three variants, referred to as I, II, and III.
A commonly adopted variant is to place the observer at $z\to\infty$
and write \Eq{eq:pol} as \citep[e.g.][]{DB90,BDMSS93,Soko98}
\EQ
P(\lambda^2)=p_0 \int_{-\infty}^\infty
\epsilon(z) e^{2\ii(\psi(z)+\phi(z)\lambda^2)}  \,\dd z
\quad\mbox{(variant I)}.
\label{pol1}
\EN
Another convenient variant is to place the observer at $z=0$
and write \Eq{eq:pol} instead as
\EQ
P(\lambda^2)=p_0 \int_{0}^\infty
\epsilon(z) e^{2\ii(\psi(z)+\phi(z)\lambda^2)}  \,\dd z
\quad\mbox{(variants II and III)}.
\label{pol2}
\EN

A second more crucial point concerns definition of the Faraday depth $\phi(z)$.
For variant I \citep[e.g.][]{DB90,BDMSS93,Soko98}, the choice is obvious
\EQ
\phi(z)=K \int_z^\infty \! n_e(s) B_\|(s) \, \dd s
\quad\mbox{(variant I)}.
\label{phi1}
\EN
However, when the observer is at $z=0$, one can define
\EQ
\phi(z)=K \int_0^z \! n_e(s) \BB\cdot\kk \, \dd s
\quad\mbox{(variants II and III)},
\label{phi2}
\EN
where $\kk$ is a unit vector pointing either toward the source \citep{Bur66}
or toward the observer \citep{Frick01}.
Thus, we have either \citep{Bur66,FSSB10,FSSB11}
\EQ
\phi(z)=K \int_0^z \! n_e(s) B_\|(s) \, \dd s
\quad\mbox{(variant II)},
\label{phi3}
\EN
or, as in the present paper and in many others \citep{Frick01,BB05,HBE09},
\EQA
\phi(z)=K \int_z^0 \! n_e(s) B_\|(s) \, \dd s
=-K \int_0^z \! n_e(s) B_\|(s) \, \dd s
\quad\mbox{(variant III)}.
\label{phi4}
\ENA
This formulation is also equivalent to the now-common notation
where one writes \citep[e.g.][]{Hea09,BHB10,Gie13}
\EQ
\phi(z)=K\int_{\rm source}^{\rm observer} n_e\BB\cdot\dd{\bm{l}}
\quad\mbox{(variant III)},
\EN
because $\BB\cdot\dd{\bm{l}}$ is the same as our $B_\|(s) \, \dd s$,
while source and observer correspond to $z$ and $0$, so the integral
goes from $z$ to $0$.

Concerning the definition of $\phi(z)$, we emphasize that
Faraday rotation of the polarization plane is a physical process that
does not depend on the coordinate system or the position of the observer.
Apparently, the sense of clockwise or counterclockwise rotation depends on the
position of the observer with respect to the polarization plane.
Consider two observers, Observer~A at $z=0$ looking in the direction of
$+\infty$ and Observer~B at $z=+\infty$ looking toward $z=0$.
The Faraday rotation corresponds then to an increase (decrease) of the
polarization angle in the $(x,y)$ plane with increasing (decreasing) $z$,
i.e., for a wave approaching Observer~B (Observer~A).
However, both observers will see counterclockwise rotation of the
polarization plane of the waves.
A common convention is that positive RM means that the line-of-sight
magnetic field between the source and the observer points toward the observer.
This is the case for \Eq{phi1} and \Eq{phi4} with
$\RM=\dd\chi/\dd\lambda^2$.
On the other hand, with \Eq{phi3} one would need to write $\RM=-\dd\chi/\dd\lambda^2$,
which is mathematically correct, but not recommended in view of
RM synthesis techniques where Faraday depth is used with the same
convention as RM.
We conclude therefore that the only meaningful definitions are
either \Eq{pol1} with \Eq{phi1} (variant~I) or \Eq{pol2} with \Eq{phi4} (variant~III).



\end{document}